# CIRCULAR MODES FOR FLAT BEAMS IN LHC*

A. Burov, FNAL, Batavia, IL 60510, USA

*Abstract*
Typically x/y optical coupling is considered as unwanted and thus suppressed; particular exclusions are electron and ionization coolers. Could some special coupled modes be effectively applied for the LHC complex? Apparently, the answer is positive: use of the circular modes [1] in the injectors with their transformation into planar modes in the LHC allows both the space charge and beam-beam luminosity limitations to be significantly reduced, if not practically eliminated.

## PLANAR AND CIRCULAR MODES

Conventional x/y betatron oscillations can be referred to as planar, since in the geometrical 3D space every one of them is seen as a plane, horizontal or vertical. Planar optical modes are described by the conventional Twiss functions; Courant-Snyder invariants and betatron phases are canonical actions and phases for these modes. In more general case, the two transverse degrees of freedom are arbitrary coupled, and their description requires more complicated set of general 4D Twiss functions, which can be taken ether in Edwards-Teng or Mais-Ripken form (see e. g. Ref. [2]). For the coupled case, there are still two Courant-Snyder invariants, expressed as quadratic forms of the 4D phase space vectors. Provided that an optical transformation is linear and symplectic, the two Courant-Snyder invariants are preserved.

An interesting example of fully coupled optics is presented by circular modes, when particles are moving along clockwise or counter-clockwise spirals [1]. These modes are eigenfunctions for rotation-invariant focusing elements, like solenoids and bending magnets with the field index ½:

$$-\frac{\partial B_y}{\partial x}\frac{\rho}{B_y} = \frac{1}{2}.$$

For quadrupole and skew-quadrupole based focusing, optical modes are generally elliptical, being more close to planar or circular in special cases. In principle, there is a direct analogy between light polarization and optics of particles in accelerators: the eigenfunctions are determined by symmetry of the media or focusing. For rotation-invariant matrices, the angular momentum is preserved; thus, the angular momentum $M = xp_y - yp_x$ has to be proportional to a difference of the two circular Courant-Snyder invariants $J_\pm$; in fact, it is just equal to that: $M = J_+ - J_-$. Since the beam emittances are nothing else as averages of the two actions, the beam angular momentum is given by a difference of its two circular emittances: $\langle M \rangle = \varepsilon_+ - \varepsilon_-$. An important aspect of the analogy between light and charged particle optics relates to planar-circular transformations. For charged particle optics, a possibility of this transformation was pointed out in Ref. [3] and practically demonstrated in Ref [4], where it was shown that these *optical adapters* can be implemented by means of skew-triplets. If a beam in a planar state is coming into a planar-to-circular adapter, the outgoing beam would be circular, and vice-versa. If a beam with very different emittances is in a planar state, it is flat. If this beam is transformed into a circular state, it becomes a round vortex, which angular momentum is equal to the larger emittance. Since adapters are based on linear optics, circular-planar transformations preserve the both emittances: $\varepsilon_\pm \leftrightarrow \varepsilon_{x,y}$.

## SPACE CHARGE SUPPRESSION

For planar modes, the space charge tune shift is determined by both emittances, being maximal for a plane of smaller emittance. That is why space charge limitation suggests for the two emittances to be close to each other. For circular modes, the space charge limitation works differently: when the emittance ratio is high, the space charge tune shift is determined by the larger emittance only – even if the smaller emittance goes to zero. Indeed, for the circular beam state, the beam size is determined by the larger emittance, making space charge insensitive to the smaller one. To make this tune shift of a circular beam with two very different emittances identical to the tune shift of a planar beam with two equal emittances, the larger circular emittance has to be two times higher than one of the planar emittances [5].

This property of the circular modes suggests an idea to use them for low-energy accelerators, with the larger emittance determined by the space charge tune shift, and the smaller one can be as small as possible. In this case, the beam brightness can be significantly increased due to reduced value of the smaller emittance. After acceleration to sufficiently high energy, the beam can be transformed into a planar state, becoming flat.

## FLAT BEAMS IN LHC

A flat beam in the collider provides significant advantages, seen in electron machines. A first one relates to the luminosity. Keeping in mind that the space charge requires larger emittance be two times higher than in the conventional planar case, it can be concluded, that luminosity can be increased as soon as the emittance ratio is 4 or higher, being proportional to the square root of the emittance ratio.

Another important benefit of the flat beam is suppression of beam-beam resonances. Indeed, a power of a resonance of an order (n,m) is determined by the beam-beam Hamiltonian term $\propto \Delta x^n \Delta y^m$. If one of the sizes goes to zero, this 2D net of the resonances reduces to 1D, allowing a lot more resonance-free space in the tune plane. As a consequence, the beam separation can be significantly reduced without any damage for emittances or lifetime.

One more important dimension opened by the flat beams is related to luminosity levelling. To do that, beta-function in the horizontal (larger emittance) plane can be increased as much as needed, gradually squeezed later with the brightness reduction. Large horizontal beta-function allows increasing the crossing angle without any luminosity reduction, similar to electron colliders. As a result, neither crab-cavity nor high aperture in the focusing triplet [6] is needed any more, and the long-range beam-beam collisions can be excluded.

Exclusion of the long-range collisions entails a valuable benefit for the coherent stability. As it is already clear from LHC operations, the octupole currents required for the stabilization are 2-3 times higher for 2 beams seeing each other compared with the single beam case. Getting rid of the LR collisions would allow working at negative octupole polarity with sufficiently low stabilizing octupole current.

## HOW TO PREPARE CIRCULAR MODES?

In some details, this problem was considered in Ref. [7]. Obviously, the beam in a circular state requires accelerator optics to be circular or reasonably close to that. To have high ratio of the emittances, the beam has to be injected predominantly into one of the circular modes, keeping another emittance as small as possible. This could be done by means of multi-turn painting with a pencil beam from a linac: the emittance ratio is determined then by a number of injection turns. Most likely, emittance ratio >10 would require space charge to be taken into account at injection: otherwise a space-charge related mismatch would heat the smaller emittance preventing having it that small. For not so high emittance ratio, like 10 or so, this mismatch, probably, can be ether neglected or compensated on average only. To inject into truly space-charge self-consistent state [8], allowing emittance ratio ~20 or higher, the inductive synchrotron can be used [9]. Application of circular modes for the LHC complex definitely requires circular optics for the Booster and PS. Is this optics required for the SPS as well, or the SPS could work with flat planar beams, is one of the opened questions.

## OPEN QUESTIONS

The described proposal of circular modes - flat beams entails a whole number of questions for the LHC complex. There preliminary list is suggested below.
1. What is the minimal lower emittance justifying application of the circular modes?
2. What is an optimal scenario for LHC for that value of the lower emittance?
3. What is luminosity lifetime for the flat beams, taken into account IBS, gas scattering, synchrotron radiation and direct burning out in collisions?
4. What can be benefits from even smaller values of the lower emittance?
5. How to minimize the lower emittance at injection from the linac?
6. Can the space charge tune shift be increased by injection into self-consistent state, as it foreseen in Ref [8]?
7. How to minimize lower emittance growth at acceleration? How to optimize optics for that?
8. Which optics has to be used at the SPS - planar or circular?
9. Can there be any use in the circular optics in the LHC? For instance – everywhere outside IRs?

These and related questions open a wide sphere of research required for any practical decision about the suggested proposal of circular modes – flat beams for the LHC.

## SUMMARY

Circular optics in the injector complex makes space charge tune shift independent of the smaller emittance, thus suggesting having it much smaller than the larger one, with significant increase of the beam brightness. This scheme requires proper multi-turn injection from the linac. Having a beam with emittance ratio ~10 or higher allows to get rid of long-range collisions in the LHC, to increase its luminosity, to reduce the aperture in the IR triplet, eliminate a need in the crab-cavity, and make the beam more stable.

Hopefully, the promising points of this concept are sufficiently attractive to start a new research program for possible future of the LHC complex, with a goal for the luminosity being as high as possible for the detectors.

## ACKNOWLEDGMENT

I am indebted to Slava Derbenev, who suggested the idea of circular modes and invented beam adapters; to Slava Danilov, with whom we started this interesting journey, and who continues to help me a lot with his advices; to Sasha Valishev, who just started to work at this concept, but already significantly contributed. My special thanks are to Elias Metral, whose inspirational interest to the circular modes gave me initial motivation to think about their possible use for the LHC.